%% file: arXiv-v1.tex
\documentclass[aps,preprint,superscriptaddress]{revtex4-2}
\usepackage{graphicx}
\usepackage{url}
\usepackage{upgreek}
\usepackage{csquotes}

\begin{document}

% Page header
%\markboth{Bowman et al.}{Remembering the work of Phillip L. Geissler}

% Title 
\title{Remembering the work of Phillip L. Geissler: A coda to his scientific trajectory}

%Authors, affiliations address.

\author{Gregory R. Bowman}
\affiliation{Bioengineering, Biochemistry and Biophysics, University of Pennsylvania, PA 19104, USA;}% email: }
\author{Stephen J. Cox} 
\affiliation{Yusuf Hamied Department of Chemistry, University of Cambridge, Lensfield Road, Cambridge CB2 1EW, UK} % ; email: sjc236@cam.ac.uk 
\author{Christoph Dellago}
\affiliation{Faculty of Physics, University of Vienna, 1090 Vienna, Austria} %; email: christoph.dellago@univie.ac.at
\author{Kateri H. DuBay}
\affiliation{Department of Chemistry, University of Virginia, Charlottesville, VA 22903, USA}
\email{dubay@virginia.edu}
\author{Joel D. Eaves} 
\affiliation{Department of Chemistry, University of Colorado Boulder, Boulder, CO 80309, USA} %; email: Joel.Eaves@colorado.edu
\author{Daniel A. Fletcher}
\affiliation{Department of Bioengineering \& Biophysics Program, University of California, Berkeley, Berkeley, United States; Division of Biological Systems \& Engineering, Lawrence Berkeley National Laboratory, Berkeley, United States; Chan Zuckerberg Biohub, San Francisco, United States}
\author{Layne B. Frechette}
\affiliation{Martin A. Fisher School of Physics, Brandeis University, Waltham, MA 02454, USA}
\email{laynefrechette@brandeis.edu}
\author{Michael Gr\"unwald}
\affiliation{Department of Chemistry, University of Utah, Salt Lake City, USA, 84112}
\author{Katherine Klymko}
\affiliation{Lawrence Berkeley National Lab, Berkeley, California 94720}
\author{JiYeon Ku} 
\affiliation{R\&D Center, Eloi Materials (EML) Co., Ltd, Suwon, Republic of Korea 16229}
\author{Ahmad K. Omar}
\affiliation{Department of Materials Science and Engineering, University of California, Berkeley, California 94720, USA; Materials Sciences Division, Lawrence Berkeley National Laboratory, Berkeley, California 94720, USA}
\author{Eran Rabani}
\affiliation{Department of Chemistry, University of California, Berkeley, California 94720, United States; Materials Sciences Division, Lawrence Berkeley National Laboratory, Berkeley, California 94720, United States; The Raymond and Beverly Sackler Center of Computational Molecular and Materials Science, Tel Aviv University, Tel Aviv 69978, Israel}
\author{David R. Reichman}
\affiliation{Department of Chemistry, Columbia University, New York, NY, USA}
\author{Julia R. Rogers}
\affiliation{Department of Systems Biology, Columbia University, New York, New York 10032, USA}
\email{jr4182@cumc.columbia.edu}
\author{Andreana M. Rosnik}
\affiliation{Atomwise, San Francisco, California 94103, USA}
\author{Grant M. Rotskoff}
\affiliation{Department of Chemistry, Stanford University, Stanford, CA 94305, USA}
\email{rotskoff@stanford.edu}
\author{Anna R. Schneider}
\affiliation{Form Energy, Berkeley, California 94710, USA}
\author{Nadine Schwierz}
\affiliation{Institute of Physics, University of Augsburg, Germany}% email: }
\author{David A. Sivak}
\affiliation{Department of Physics, Simon Fraser University, Burnaby, BC, V5A1S6, Canada}
\email{dsivak@sfu.ca}
\author{Suriyanarayanan Vaikuntanathan}
\affiliation{Department of Chemistry, University of Chicago, Chicago, Illinois 60637, USA} %; email: svaikunt@uchicago.edu
\author{Stephen Whitelam}
\affiliation{Molecular Foundry, Lawrence Berkeley National Laboraratory, Berkeley, CA 94709, USA}
\email{swhitelam@lbl.gov}
\author{Asaph Widmer-Cooper}
\affiliation{School of Chemistry, The University of Sydney, Sydney, 2006 Australia}%; email: asaph.widmer-cooper@sydney.edu.au

%Greg Bowman, 

% Stephen J. Cox

% Christoph Dellago, 

%Kateri H. DuBay, 

% Joel D. Eaves, 

% % Daniel A. Fletcher

%Layne B. Frechette, 

%Michael Gruenwald, 

% katie klymko

%JiYeon Ku, 

% amhad omar

%Eran Rabani

% David R. Reichman

% Julia R. Rogers

% Andreana M. Rosnik

%Grant M. Rotskoff, 

% Anna R. Schneider

%Nadine Schwierz, 

% David A. Sivak

% Suriyanarayanan Vaikuntanathan, 

%Stephen Whitelam, 

% Asaph Widmer-Cooper, 

%Abstract
\begin{abstract}
Phillip L. Geissler made important contributions to the statistical mechanics of biological polymers, heterogeneous materials, and chemical dynamics in aqueous environments. He devised analytical and computational methods that revealed the underlying organization of complex systems at the frontiers of biology, chemistry, and materials science. In this retrospective, we celebrate his work at these frontiers.
\end{abstract}

%Keywords, etc.
% \begin{keywords}
% statistical mechanics, chemical dynamics, biological systems, aqueous environments, algorithm development, model development
% \end{keywords}

\maketitle

%Table of Contents
\tableofcontents

\section{Preface}
Phillip Geissler was a valued member of the editorial board for Annual Reviews of Physical Chemistry since 2012. We were looking forward to seeing him again at the 2022 editorial board meeting, especially after two years of virtual meetings due to COVID-19. It was devastating to hear that Phill had been taken from us this summer. Phill was a champion of young faculty and a fount of good advice. His perennial good judgment strengthened Annual Reviews of Physical Chemistry and the editorial board will miss him dearly. 
 
Speaking personally, Phill and I shared a birthday week, which gave us at least one excuse annually to celebrate (or more likely commiserate!) together. We ribbed each other about the relative merits of not only ``Team Quantum'' vs. ``Team StatMech,'' but also Cal vs. Stanford. He chuckled at my attempts to learn the guitar, but he assured me (with a twinkle in his eye) that I should take his bemusement as encouragement. I was delighted (but not at all surprised, since Phill was much loved) that many past Geissler group members were excited to take on the task of writing a summary of Phill's many contributions to theoretical chemistry. Phill was an amazing theorist, a generous mentor, and a gifted teacher (for both undergraduates and graduate students). I think this chapter captures the breadth and depth of Phill's contributions much better than I ever could and thank the authors for their work and dedication. I will miss Phill and our field is poorer for his loss.    

-Todd Martinez

\section{Introduction by the Geissler Group}

One of the challenges we faced with this retrospective on the work of Phill Geissler is the sneaking suspicion that however much we polished the manuscript, Phill might consider the work to be a promising first draft on the way to a solid second. Phill was a scholar and a leading figure in the field of statistical mechanics, but he was also a wordsmith and a poet. He chose his words carefully, not content to convey results when he could also convey the ideas and concepts that underpinned them. He delighted in choosing talk titles that were both playful and deep. His talks, speaker introductions, and annual state-of-the-group meetings could at times rise to the level of oratory. So it is natural to introduce this retrospective with Phill's summary that 

\begin{displayquote}
The Geissler Research Group focuses on the statistical mechanics of biological polymers, of heterogeneous materials, and of chemical dynamics in aqueous environments. Although these topics are physically diverse, they are unified by features of disorder and strong non-covalent interactions among many molecules. As such, they are amenable to similar approaches and can sometimes be understood in common terms. Exploiting this connection, the group devises analytical and computational methods to reveal the underlying organization of complex systems at the frontiers of biology, chemistry, and material science.
\end{displayquote}

In this volume we explore Phill's work at these frontiers. We cover water (Sec.~\ref{sec:water}); biophysical systems (Sec.~\ref{sec:biophysics}); self-assembly (Sec.~\ref{sec:selfassembly}); nanomaterials (Sec.~\ref{sec:nanomaterials}); and model- and algorithm development (Sec.~\ref{sec:methods}). In each section, the overarching themes of Phill's work are evident: his ability to choose important and rewarding problems; his focus on the fundamentals and on ``identifying the essential microscopic variables whose fluctuations cannot be ignored''; his fascination with the subtle as well as the simple; and the inspiration he took from experiment and the collaborations with experimentalists that were central to his career. 

Phill was a brilliant and creative scientist. He had high standards, and demanded the same from his group. He was also genuine and humble, generous with his advice and encouragement, and he liked a good laugh. We cannot summarize his work as he would have done, but it is our privilege to try. This retrospective is our tribute to Phill, our teacher, mentor, and friend.

\input{water.tex}

\input{biophysics.tex}

\input{selfassembly.tex}

\input{nanomaterials.tex}

\input{methods.tex}

\section{The trajectory from here}

We hope that presenting Phill's collected contributions in a single volume sheds light on the underlying themes that informed his scientific work. 
Dynamical trajectories shaped his understanding of physical systems, and his research continually emphasized the necessity of accounting for the collective fluctuations characteristic of the nanoscale.
He sought to explain experimental observations in exceedingly complex systems by devising models that captured the essential fluctuations and nothing more. 
His approach to research, like his approach to teaching, was guided by an appreciation for clarity, simplicity, and elegance.
While the physical systems he studied were not constrained by disciplinary boundaries, he found a common language to explain complex processes from ion solvation to biological self-organization.
Phill will remain to us a model of a scientist, the one who showed us the ropes, and a friend whom we will miss immensely.
We hope that this perspective remembering his brilliant and too-short career will light the path as we navigate the unknown and rugged landscapes ``at the frontiers of biology, chemistry, and materials science.''

% %Disclosure
\section*{DISCLOSURE STATEMENT}
The authors are not aware of any affiliations, memberships, funding, or financial holdings that might be perceived as affecting the objectivity of this review. 

% %Contributions
\section*{AUTHORS CONTRIBUTIONS STATEMENT}
K.H.D., L.B.F., J.R.R., G.M.R., D.A.S., and S.W. conceived and designed the review article.
S.J.C., J.D.E., G.M.R., and S.V. wrote Sec.~\ref{sec:water}.
G.R.B, K.H.D., D.A.F., J.R.R., A.M.R., A.R.S., N.S., D.A.S., and S.W. wrote Sec.~\ref{sec:biophysics}.
M.G., K.K., J.K., A.O., E.R., D.R., G.M.R., S.W., and A.W. wrote Sec.~\ref{sec:selfassembly}.
L.B.F., M.G., and A.W. wrote Sec.~\ref{sec:nanomaterials}.
C.D., G.M.R., and S.V. wrote Sec.~\ref{sec:methods}.
All authors reviewed and edited the article.

% % Acknowledgements
\section*{ACKNOWLEDGMENTS}

This retrospective is the product of contributions from many of Phill's students and collaborators, but the work that is summarized would never have been possible without the members of the Geissler group throughout the years. So, with that in mind, we would like to acknowledge everyone: 
Adrianne Zhong,
Ahmad Omar,
Amr Dodin,
Andrea Pasqua, 
Andreana Rosnik, 
Anna Schneider, 
Arpita Mandan, 
Asaph Widmer-Cooper, 
Brad Compton, 
Brian Gin, 
Carl Rogers, 
Chris Ryan, 
David Moler, 
David Sivak, 
Dayton Thorpe, 
Easun Arunachalam, 
Eion Laighleis, 
Evan Hohlfeld, 
Evan Wang, 
Georg Menzl, 
Grant Rotskoff,
Gustavo Espinoza Garcia,
Jaffar Hasnain, 
Jiyeon Ku, 
John Haberstroh, 
Joseph Harder, 
Joyce Noah-Vanhoucke, 
Julia Rogers, 
Julian Weichsel, 
Kateri DuBay,
Katherine Delevaux,
Katie Klymko, 
Katie Martins,
Kritanjan Polley,
Laura Armstrong, 
Layne Frechette, 
Lisa Littlejohn,
Lucie Liu, 
Lutz Maibaum, 
Michael Gruenwald, 
Nadine Schwierz, 
Nathan Odendahl, 
Patrick Shaffer, 
Pratima Satish,
Paul Wrona,
Ramin Khajeh, 
Rian Kormos,
Sam Oaks-Leaf,
Sander Pronk, 
Sean Cray, 
Steve Cox, 
Steve Whitelam, 
Sucheol Shin, 
Suri Vaikuntanathan, 
Todd Gingrich,
Ty Perez,
Wei Zhang, 
Will Browne, 
Yizhi Shen. 

%Acknowledgements, general annotations, funding.
J.R.R. acknowledges a Fellowship of The Jane Coffin Childs Memorial Fund for Medical Research. S.J.C is funded by a Royal Society University Research Fellowship (URF\textbackslash R1\textbackslash 211144). S.W. performed work at the Molecular Foundry, Lawrence Berkeley National Laboratory, supported by the Office of Science, Office of Basic Energy Sciences, of the U.S. Department of Energy under Contract No. DE-AC02--05CH11231. G.M.R. performed work supported by the U.S. Department of Energy, Office of Science, Office of Basic Energy Sciences, under Award Number DE-SC0022917.

%\nocite{*}
\bibliographystyle{ar-style3.bst}
\bibliography{geissler_collected_papers,non_geissler,assembly_bib,nano}

\end{document}

%% file: water.tex
\section{Water} \label{sec:water}

% Main authors: Grant and Stephen Cox
% Contributing: Joel, Suri
% did not have contact info for Joyce

%Authors: Grant Rotskoff, Joel Eaves, Suriyanarayanan Vaikuntanathan, Stephen Cox

Water, especially in its liquid state, remains a surprising and intricate puzzle for physical chemistry. In Phill’s own words, “water is a famously unusual liquid”~\cite{geissler_water_2013}, an eccentricity inherited from its strongly directional interactions and complex, but persistent, hydrogen bond structure~\cite{eaves_hydrogen_2005}. 
%In some sense, 
%suggest removing "in some sense"
%as more categorically “traditional” physical chemistry than many of the other topics that he studied. 
Phill’s work on water brought to the study of aqueous solutions the same insight, creativity, and interdisciplinary perspective that he applied to his work more broadly. %though the topic is traditional physical chemistry.
%"that he applied to his work more broadly"? Might work better with the following sentence
% drove him to expand the boundaries of statistical mechanics to such a diverse array of topics. 
Judicious use of transition path sampling, targeted minimal models, and clear statistical mechanical analysis of complicated experimental measurements all permeate his work in this domain. 

\begin{figure}
    \centering
    \includegraphics[width=\linewidth]{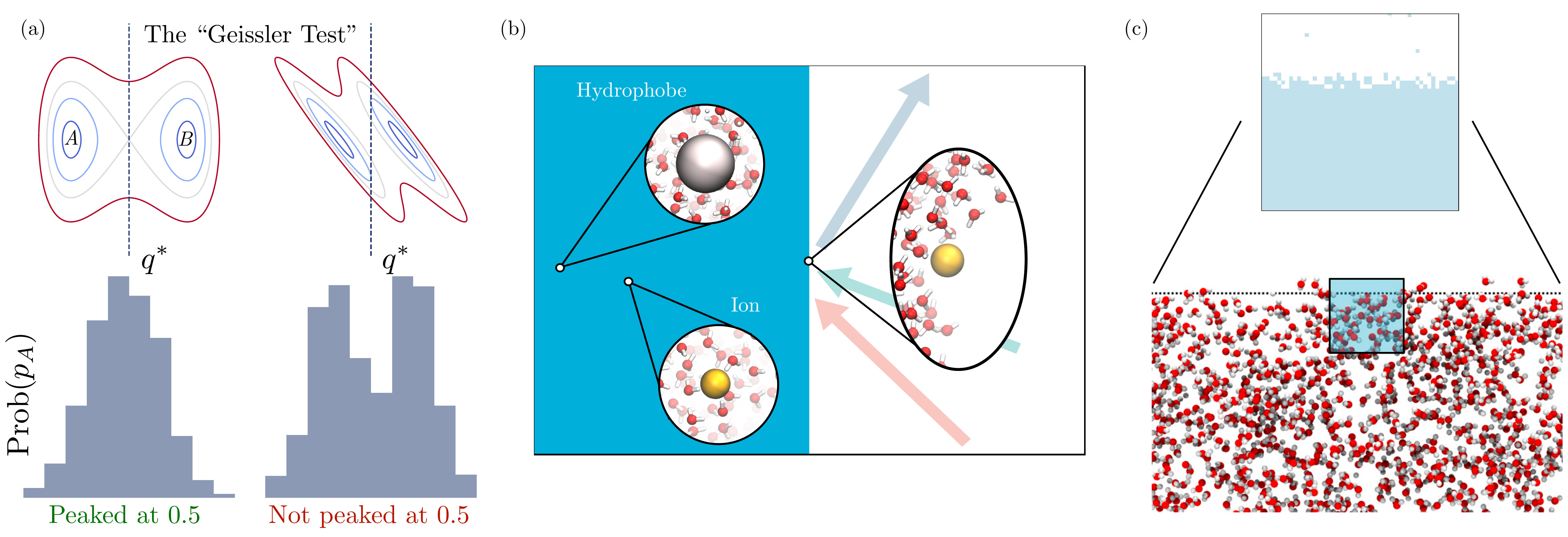}
    \caption{Phill's work on liquid water focused on the subtle fluctuations that dictated kinetics, solvation, and ion-specific effects. His approach was deeply influenced by transition path sampling and the limitations of local reaction coordinates, as illustrated by the Geissler test in which the quality of reaction coordinate is examined by estimating the committor distribution from the top of the free energy barrier (a); a good reaction coordinate has a distribution that is peaked at $p_A = 0.5$, meaning half of all trajectories react. Solvation and surface specific effects were probed with a variety of spectroscopic methods, illustrated in (b), and Phill's approach to both hydrophobicity and ion solvation were heavily influenced by Lum-Chandler-Weeks theory, which yielded quantitatively accurate lattice models, depicted schematically in (c). }
    \label{fig:water}
\end{figure}

Phill’s longest-running research project was water: In one of his first papers as a Ph.D. student in David Chandler’s group at Berkeley, Phill applied the recently developed transition path sampling method~\cite{bolhuis_transition_2000}, described in Sec.~\ref{sec:methods}, to understand the dissociation of a Na$^+$Cl$^-$ ion pair. 
At the time, estimates of the dissociation rate based on transition state theory were an order of magnitude too small. Phill’s work diagnosed the issue by showing that an ionic separation distance was insufficient as an order parameter to describe the kinetics of the process. Sampling trajectories, on the other hand, enabled him to identify the neglected, rare solvent fluctuations that ultimately dictated the rate of dissociation. Phill always insisted on careful consideration of appropriate order parameters and the first figure of this early paper was an image he would go on to draw for students time and time again to illustrate the dangers of a bad reaction coordinate (Fig.~\ref{fig:water} a). His paper also introduced a diagnostic that became a standard test in the Geissler group and beyond: “committor analysis” proceeds by sampling trajectories from the top of a free energy barrier to ensure that the order parameter truly discriminates between metastable basins (Fig.~\ref{fig:water} a). This analysis was affectionately dubbed the “Geissler test” in the Geissler group.

Technical innovations in path sampling enabled a more complete understanding of the complex and collective fluctuations that drove rare processes in aqueous solutions, and appreciating the importance of rare solvent fluctuations for dynamics in the condensed phase became a central theme of Phill’s work. Much of his subsequent thesis work focused on the dynamics of proton transfer in liquid water, a fundamental problem that underlies acid-base chemistry. Working with Michele Parrinello’s group in two separate trips to the Max Planck Institute in Stuttgart, Germany, Phill combined transition path sampling methods with Car-Parrinello molecular dynamics to study proton transfer, initially in a protonated water trimer~\cite{geissler_ab_2000}, a system he had previously studied using empirical models~\cite{geissler_chemical_1999}. This work culminated in a landmark paper published in the journal \emph{Science} in which Phill and collaborators from the Chandler and Parrinello groups demonstrated that autoionization in liquid water is driven not only by rare fluctuations in the solvent electric field that cleave an oxygen-hydrogen, but also by coincident reorganization of the hydrogen-bond wire,
an event that prohibits recombination~\cite{geissler_autoionization_2001}. By harvesting reactive trajectories, this paper clearly demonstrated the fundamental inadequacy of a local order parameter for autoionization.
%demonstrated by harvesting trajectories that local order param not adequate?
These calculations highlight many of the features that would go on to influence Phill’s perspective on liquid water, namely the importance of the hydrogen-bond network, the influence of rare electric field fluctuations for ion solvation~\cite{geissler_importance_2000}, and the necessity of carefully attending to the collective fluctuations that dictate kinetics in aqueous solutions.

The central importance of water’s hydrogen bonding network and its implications for spectroscopic measurements re-emerged in Phill’s collaborations with Andrei Tokmakoff and Rich Saykally.
Phill often joked about his disdain for quantum mechanics, though it is somehow not surprising that he made critical contributions to the theory and interpretation of the vibrational spectra of water. In the early 2000s, experimentalists were developing sophisticated nonlinear optical methods to probe liquid motions on sub-picosecond time scales, and they turned their attention to water. Badger’s Rule, an empirical law from steady-state spectroscopy, states that the frequency of the OH stretch shifts to the red with increasing hydrogen bond strength. Phill’s work on infrared photon echo and two-dimensional spectroscopies showed how time-dependent frequency shifts of the OH stretch probe the evolution of molecular structure in water~\cite{fecko_ultrafast_2003, eaves_electric_2005, eaves_hydrogen_2005}. While some features in those spectra had simple molecular descriptions, others did not. Phill showed that they were related to collective rearrangements of the liquid that result when molecules “switch allegiances” between their hydrogen bond partners. The short lifetime of a putative broken hydrogen bond in liquid water at ambient temperatures shows that these bonds are broken, but only fleetingly~\cite{eaves_hydrogen_2005}.

When working with his group, Phill loved to examine dynamical trajectories of the models being developed. An appreciation for the complexity of molecular relaxation clearly motivated his thinking about the interpretation of spectroscopic measurements. For example, while many had interpreted Raman spectra of liquid water to indicate that two distinct classes of hydrogen bonds existed, dynamical trajectories of model systems led Phill to conclude that any attempt to distinguish between two such classes was ultimately arbitrary~\cite{smith_unified_2005}. Moreover, while the existence of an 
isosbestic point in the Raman and IR spectra of liquid water had been interpreted as evidence of two interconverting species, Phill provided an elegant and minimal argument that this interpretation was wrong~\cite{geissler_temperature_2005}, and that the isosbestic point was simply an indication of an order parameter that was insensitive to changes in the temperature over the range probed by the experiment. In fact, he showed that isosbestic points can arise even in a thermal distribution of harmonic oscillators~\cite{geissler_temperature_2005}, and Monte Carlo simulations of water demonstrate that the OH bond distance in water is nearly temperature independent at the isosbestic point of the Raman spectrum, despite the fact that the distribution has only one dominant state~\cite{smith_unified_2005}.

The implications of a robust hydrogen bond network for ion solvation~\cite{smith_effects_2007} and interfacial properties subsequently became a major thrust of Phill’s work on water. Simulations of air-water interfaces were a crucial tool to inform his thinking, in part due to the subtlety of the indirect information reported by surface measurements such as sum frequency generation (SFG)~\cite{noah-vanhoucke_toward_2009,noah-vanhoucke_statistical_2009}. Phill and coworkers sought to elucidate the microscopic origin of hyperpolarizability of the air-water interface by stripping down the measures of orientational bias to just OH and OD bond vectors, a vast simplification compared to existing approaches. By reducing the complexity, it became possible to diagnose the effects of various ions on the SFG spectra, ruling out local effects on solvation structure which were largely spherically symmetric and hence undetectable with SFG.
In recent work, Phill showed that ice-like local structure exists at air-water interfaces, yet another manifestion of interfacial effects on the hydrogen bond network~\cite{odendahl_local_2022}.
Many aspects of this problem continued to occupy his work, including several studies exploring ion-specific effects as codified by the Hofmeister series. 

Phill’s work on solvation was heavily influenced by the seminal work by Ka Lum, David Chandler and John Weeks \cite{lum_hydrophobicity_1999}, which showed how the forces driving hydrophobic assembly can be quantitatively captured in terms of a framework that resolves the fluctuations on both short and long length scales (Fig.~\ref{fig:water}). While the eponymous LCW theory was indeed a remarkable advance, the resulting theoretical framework still required a fitting parameter with an unclear physical meaning. In part motivated by his work on ion solvation, Phill and his co-workers recognized that the LCW framework did not completely account for the effect of the low energy capillary modes. By including the physics of rough capillary waves in the LCW framework they were able to construct a quantitatively accurate theory for hydrophobic solvation without the aid of any fitting parameters \cite{vaikuntanathan_putting_2014,vaikuntanathan_necessity_2016}. The theory only took as input the physics of short length scale fluctuations, as parameterized by the oxygen-oxygen two-point correlations in the bulk, and the physics of long length scale fluctuations as parameterized by the surface tension of water. With just these inputs, this LCW-inspired theory predicted the free energies of hydrophobic solvation across a large range of sizes and shapes.

Phill strove to understand ion solvation in a similar vein, that is, to faithfully account for solvent fluctuations and their modification by the solute. Phill’s clear statistical mechanical analysis of ion solvation resulted in insights that challenged the prevailing understanding in the field. Consider for example the driving forces that govern the relative stability of ions at the liquid-vapor interface. A conventional accounting of the driving forces would lead us to expect an entropically favored force driving the ion from the bulk to the free interface, and an energetic (or enthalpic) driving force that keeps the ions solvated. Phill correctly recognized that this accounting missed contributions from low-energy fluctuations that populate the interface \cite{noah-vanhoucke_fluctuations_2009, otten_elucidating_2012, mccaffrey_mechanism_2017}. The effects of these fluctuations, commonly referred to as “capillary wave fluctuations,” are most pronounced on long length scales. Phill and his co-workers provided detailed and clear statistical mechanical analysis to show how these low-energy, long-wavelength modes modify the driving forces for ion solvation in counterintuitive ways. In particular, they make it entropically unfavorable for an ion to migrate to the interface. Their statistical mechanical analysis also showed that enthalpic forces drive ion solvation at interfaces. Phill and his coworkers were able to obtain analytical expressions for these forces and the resultant free energies by constructing a lattice-based model \cite{vaikuntanathan_adsorption_2013}. 

The softness of the air-water interface, essentially its ability to ``wrap around'' small ions and make their local environments similar to those of bulk, is a feature inherently beyond the scope of approaches rooted in simple dielectric continuum theory. Several of Phill’s later works therefore focused on ways to go beyond dielectric continuum theory
(DCT) and, more generally, linear-response approximations. In particular, the origin of ``charge asymmetry'', that is, the difference in solvation behavior of solutes that only differ in the sign of their charge, was a problem that Phill was determined to frame in terms of solvent fluctuations. Inspired by similar ambiguities encountered in trying to assign water molecules to “bulk” or “interface” when computing SFG spectra~\cite{noah-vanhoucke_toward_2009,byrnes_ambiguities_2011}, Phill was keen to emphasize that ion adsorption to the air-water interface cannot be understood simply by considering contributions to the electrostatic potential felt by a solute that arise from the macroscopic interface. Through careful analysis, Phill showed that for small solutes, nonlinear contributions from local solvent rearrangements dominate the solvation process \cite{cox_assessing_2020}. While the problems with DCT at the air-water interface are relatively easy to assess, developing a theoretical framework \`{a} la LCW is significantly more challenging. Nonetheless, Phill took strides toward such a field-theoretical perspective in the context of bulk ion solvation by considering symmetry constraints placed on a water molecule’s quadrupole in relation to its dipole \cite{cox_quadrupole-mediated_2021}. The resulting field theory incorporated charge asymmetry as an emergent phenomenon, while preserving the simplicity of DCT. The approach was typical of Phill: To first consider the problem in all of its technical complexity, and then, with a few clearly stated approximations, arrive at a simple result. 

Although Phill endeavored to go beyond DCT, his later work also clarified instances where it reasonably describes water’s polarization fluctuations. For example, Phill used DCT to understand how computed solvation free energies tend toward the thermodynamic limit \cite{cox_interfacial_2018}. 
Not only did this ensure that his analyses and theories were based on sound physical principles, but he was also able to conclude that water behaves like a simple dielectric medium down to nanometer length scales.%, a conclusion bolstered by subsequent work showing that .
His final contributions were to use this approach to help understand water's dielectric response under confinement \cite{cox_dielectric_2022}, and in forthcoming work, to show that even in regions close to the interface, polarization fluctuations are consistent with DCT, all the way down to microscopic probe volumes. 

%% file: biophysics.tex
\section{Biophysics}\label{sec:biophysics}

% \begin{itemize}
% \item Main Authors: Kateri DuBay, David Sivak, Julia Rogers
% % , Grant Rotskoff
% % GR: I played a small role in organizing, but you all covered everything!

% \item Contributing Authors: 
% Steve Whitelam, Dan Fletcher, Greg Bowman, Nadine Schwierz, Andreana Rosnik, Anna Schneider
% \end{itemize}

%\begin{itemize}
%\item (David contacted) Steve Whitelam, Dan Fletcher, (Eugene Shakhnovich responded but didn't provide content)
%\item (Kateri contacted) Greg Bowman, Nadine Schwierz, (Brian Gin no response)
%\item (Julia contacted) Andreana Rosnik, Anna Schneider
%\end{itemize}

%Current thematic focus:

%\begin{enumerate}
%    \item Bridging molecular and macroscopic phenomena across length and timescales
 %   \begin{enumerate}
  %      \item in terms of the types of biological processes investigated
   %     \item In terms of the methods used to study them: Elegant minimal model development and sampling methodology
    %\end{enumerate}
    %\item Collective effects, fluctuation induced phenomena, and otherwise rare events – including role in spatial organization, phase separation
    %\item Nonequilibrium effects and dynamics – may want to divide between general kinetics and active processes
    %\item Making direct comparisons to experiment, often through minimalist biological experiments (reconstitution)
%\end{enumerate}

Biological systems feature a rich interplay of molecular and macroscopic events happening from femtoseconds to millions of years. Phill was fascinated by the coupling of such disparate scales and driven to understand such behavior in intuitive physical terms. To explain incredibly complex biological phenomena, Phill masterfully crafted surprisingly simple phenomenological models. He frequently pressed to use fewer assumptions, simpler functional forms, fewer fitting parameters, less ``magic''---the only ``magic'' should be the beauty of the emergent phenomenon. Not only aesthetically pleasing, this philosophy led to coarse-grained models whose computational tractability was key to probing length and time scales relevant to experiment. By judiciously combining this approach with atomistically detailed models, Phill and his group uncovered the microscopic events essential for triggering collective responses in myriad biological systems.

%Subsection: Biopolymers

\begin{figure}[]
   \centering
   \includegraphics[width=\linewidth]{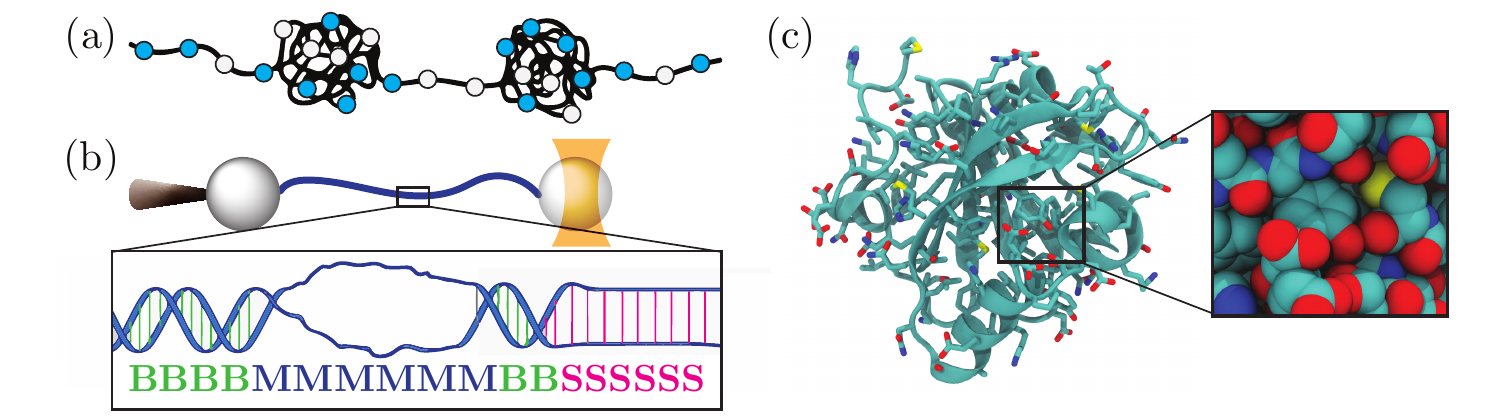} 
   \caption{Phill highlighted the importance of fluctuations in biological phenomena, such as in his studies of (a) random heteropolymer structures~\cite{geissler_solvation_2004,geissler_reversible_2002,geissler_mechanical_2002}, (b) nonequilibrium dynamics of DNA stretching~\cite{whitelam_there_2008,whitelam_stretching_2008,whitelam_microscopic_2010}, and (c) conformational rearrangements of protein side-chains in crowded environments~\cite{dubay_calculation_2009,dubay_long-range_2011,bowman_equilibrium_2012,bowman_extensive_2014, dubay_fluctuations_2015}.}
   \label{fig:polymer_prot}
\end{figure}

Phill’s first forays into biophysics took inspiration from his research on water. With his postdoctoral advisor Eugene Shakhnovich, Phill studied the role of solvent interactions in the mechanical behavior of random heteropolymers, simple models for proteins (Fig.~\ref{fig:polymer_prot}(a)). Using linear-response theory applied to a random energy model of surface monomer conformations, he showed the importance of fluctuations in tempering the predominance of ``solvophilic’’ monomers (those with strong affinity for solvent) at the surface~\cite{geissler_solvation_2004}. He also used replica mean-field theory to show that when such random heteropolymers are stretched, the variation in solvophilicity along the polymer produces partially unfolded ``necklacelike’’ structures (with compact solvophobic stretches and extended solvophilic stretches) at intermediate pulling forces, thus broadening the otherwise-sharp coil-globule transition, with important implications for mechanical strength~\cite{geissler_reversible_2002,geissler_mechanical_2002}.

Work on minimal polymer models under mechanical stress naturally led Phill to experiments from Carlos Bustamante's group showing that DNA, when pulled along its axis, abruptly extends or ``overstretches'' by about 70\% at a force of about 65 pN~\cite{smith1996overstretching}. Notably, the kinetics of overstretching vary by conditions. At low temperatures or high salt concentrations, overstretching and the subsequent shortening of the molecule occur in a reversible way, with force-extension curves superposing. But at high temperatures or low salt concentrations, the stretching-shortening cycle is hysteretic~\cite{mao2005temperature}. Two competing pictures of this process had been put forward in the literature. One hypothesis assumed that extended DNA was melted, with the base pairing between DNA strands disrupted~\cite{wenner2002salt}. A second picture argued that the mechanics and thermodynamics of overstretching could only be explained if there existed a base-paired, elongated state of the molecule called S-DNA~\cite{cocco2004overstretching}. Phill and his group developed a simplified lattice model of DNA under tension, in which base pairs adopted discrete states~\cite{whitelam_there_2008,whitelam_stretching_2008,whitelam_microscopic_2010} (Fig.~\ref{fig:polymer_prot}(b)). Dynamical simulations reproduced the condition-dependent hysteresis seen in experiment, but only if the model included the possibility of forming S-DNA. Otherwise, overstretching within the model always involved base-pair disruption and the hysteresis associated with the slow reattachment of the two DNA strands. This work showed that the nonequilibrium dynamics of single-biomolecule manipulation studies could discriminate between competing microscopic theories of the resulting structural transitions, in this case providing clear support for the S-DNA hypothesis. Subsequent experiments provided direct evidence for the existence of S-DNA~\cite{fu2010two,zhang2012two}.

%The example of S-DNA illustrates that Phill was a master of the art of crafting a surprisingly simple phenomenological model (such as a lattice model) that is able to recapitulate surprisingly complex properties of a nanoscale system (such as a DNA strand). Phill was also a master at teaching this approach. The word ``phenomenological'', a tongue-twister for some, came up so often in conversation with Phill that it would roll off the tongue mellifluously after a period of working with him. He frequently pressed to use fewer assumptions, simpler functional forms, fewer fitting parameters, less ``magic''---the only ``magic'' should be the beauty of the emergent phenomenon. In addition to appreciating the aesthetics of coarse-grained phenomenological models, Phill was also drawn to them for practical scientific reasons: they are computationally tractable when probing length- and time-scales relevant to experiment, and they lend themselves well to testing crisply tailored hypotheses. Phenomenological models became one of the key threads that tied together his forays into highly disparate areas of biophysics and beyond.

Phill continued to be interested in modeling the mechanics and dynamics of DNA as it has important implications for cellular processes (such as chromosomal compaction and segregation, viral packaging, and transcriptional regulation) that involve sharply bending DNA in a controlled fashion~\cite{Garcia:2007hi}. The wormlike chain model predicts that DNA reacts to bending stresses by deforming uniformly along its contour. However, when applied forces become very large, or equivalently during large thermal fluctuations, this deformation may be concentrated in localized excitations that render short stretches of the chain (``melts'') very pliable. Phill and his group developed coarse-grained DNA models incorporating such melts and showed their significant impact on mechanical properties that are sensitive to rare fluctuations, as probed by several experimental approaches: threading through nanopores~\cite{trepagnier_controlling_2007}; cyclization kinetics~\cite{sivak_consequences_2012}; FRET~\cite{shroff_optical_2008}; and SAXS~\cite{mastroianni_probing_2009}.

%Subsubsection: Protein Work
DNA was not the only biological polymer whose dynamics Phill sought to better understand. With coworkers, he extended early lattice protein models to investigate the folding dynamics of hundreds of thousands of heteropolymer sequences that folded to a well-ordered globular structure using a G$\bar{\rm o}$-like model \cite{Go_1983} with heterogeneous contact energies \cite{gin_limited_2009}. The appearance order of the native state’s close contacts during folding remained remarkably invariant to the removal of nonnative interactions, although the folding time scales shifted, particularly for the slower-folding sequences.
This insensitivity in the folding mechanism to nonnative interactions has been verified in subsequent simulations on both lattice polymers \cite{Faisca_nonnative_2010} and all-atom MD simulations of actual fast-folding proteins \cite{Best_PNAS_2013, Lindorff-Larsen_folding_2011}.
Further work by Phill and collaborators demonstrated that heterogeneity among the contact interactions grants unique dynamical properties to the folding trajectories \cite{mey_rare-event_2014}.

Once proteins have folded to their native state, their conformational fluctuations are greatly diminished, leading to the general perception of a relatively static native fold. However, substantial side-chain rearrangements remain sterically accessible \cite{Kussell_side-chain_2001},
prompting Phill and coworkers to probe these more subtle side-chain dynamics using Monte Carlo (MC) simulations of side-chain rotations on a natively folded and fixed backbone (Fig.~\ref{fig:polymer_prot}(c)) \cite{dubay_calculation_2009}. This simplified model enabled the quantification of side-chain entropy within the native state. By observing the range of side-chain fluctuations across a series of well-folded globular proteins, the entropic contribution available to regulate the free energy of ligand binding or protein-protein interactions from this reservoir was found to be sizable \cite{dubay_calculation_2009}. These results supported accumulating evidence from NMR order parameters attesting to the importance of side-chain entropy in regulating protein thermodynamics \cite{Frederick_Nature_2007} and were able to explain the different binding entropies between calmodulin and a series of ligands that had been previously measured by isothermal calorimetry \cite{Frederick_Nature_2007}. 
Recent work has provided more direct experimental evidence of this regulation in action by measuring differences in the conformational heterogeneity of side-chains upon ligand binding across several hundred crystallographic datasets of paired unbound and ligand-bound structures \cite{Fraser_crys-apo-holo_2022}.

To investigate the native-state fluctuations further, Phill and coworkers constructed Markov State Models (MSMs) from extensive MD simulations, which confirmed the presence of side-chain dynamics in protein cores and provided evidence of long time scale backbone dynamics \cite{bowman_extensive_2014}. Exciting functional phenomena were also observed, including the transient formation of cryptic binding pockets and allosteric communication between distant parts of the protein \cite{bowman_equilibrium_2012}. Both the MSM models and the earlier MC simulations showed that allosteric signals can be transmitted across long distances even in the absence of significant backbone motion \cite{bowman_equilibrium_2012, dubay_long-range_2011}. 
Subsequent work built on this foundation has focused on understanding and exploiting protein conformational heterogeneity in areas ranging from COVID-19 to Alzheimer's disease, in one case uncovering hidden allosteric sites in TEM-1 beta-lactamase, an important antibiotic target \cite{Bowman_Marqusee_2015}.

Proteins not only fold into native states, but also into assemblies of misfolded structures, such as the filamentous aggregates of A${\upbeta}$-peptides that are a hallmark of Alzheimer’s disease. Understanding the molecular pathway of peptide assembly and fibril growth is of great biomedical importance but has proven computationally challenging due to the long time scales involved. By extracting free energy and diffusion profiles from extensive all-atom simulations, Phill and coworkers highlighted the importance of solvation entropy and collective water rearrangements on the molecular pathways of A${\upbeta}$-amyloid fibril growth by elongation, fragmentation \cite{schwierz_dynamics_2016}, and surface-activated secondary nucleation \cite{schwierz__2017}.

%Subsubsection: Cytoskeletal filaments and their interactions with biological membranes
Phill's interest in the rich behavior that emerges from fluctuations in complex systems found a natural home in studying the principles underpinning assembly and organization of collections of diverse biomolecules. Phill applied characteristically simple models to explain puzzling results from the lab of his friend and longstanding collaborator, Dan Fletcher: in \emph{in vitro} reconstitutions involving lipid membranes and actin filaments, actin filaments polymerized on the surface of deformable lipid vesicles resulted in the formation of long filopodia-like structures~\cite{liu_membrane-induced_2008,pronk_limits_2008}. This was unexpected in two ways: first, polymerization of individual actin filaments cannot generate sufficient forces to deform planar membranes into tubes, and second, filaments beyond a certain length were expected to buckle under the restoring force of the membrane. Models from Phill's group, together with experiments from Dan Fletcher's lab, demonstrated that a deforming membrane could couple multiple actin filaments that together could overcome the barrier to tube formation~\cite{liu_membrane-induced_2008} and that filaments contained within a membrane tube do not experience conventional Euler buckling because of how the restoring force is applied~\cite{pronk_limits_2008}. Later \emph{in vitro} experiments with curved actin filaments and the side-binding protein Arp2/3 revealed a bias to bind to the outside of the filament curve, rather than the inside~\cite{risca_actin_2012}. The bending energy associated with the filament was insufficient to explain the results, so Phill devised a fluctuation-based `gating’ model that captured the bias. This view of biological materials as active, based on their assembly and disassembly dynamics in a thermally driven environment, provides a framework that continues to be relevant to biophysical problems today \cite{fletcher_active_2009}.

%Subsection: Biological membranes

\begin{figure}[]
   \centering
   \includegraphics[width=\linewidth]{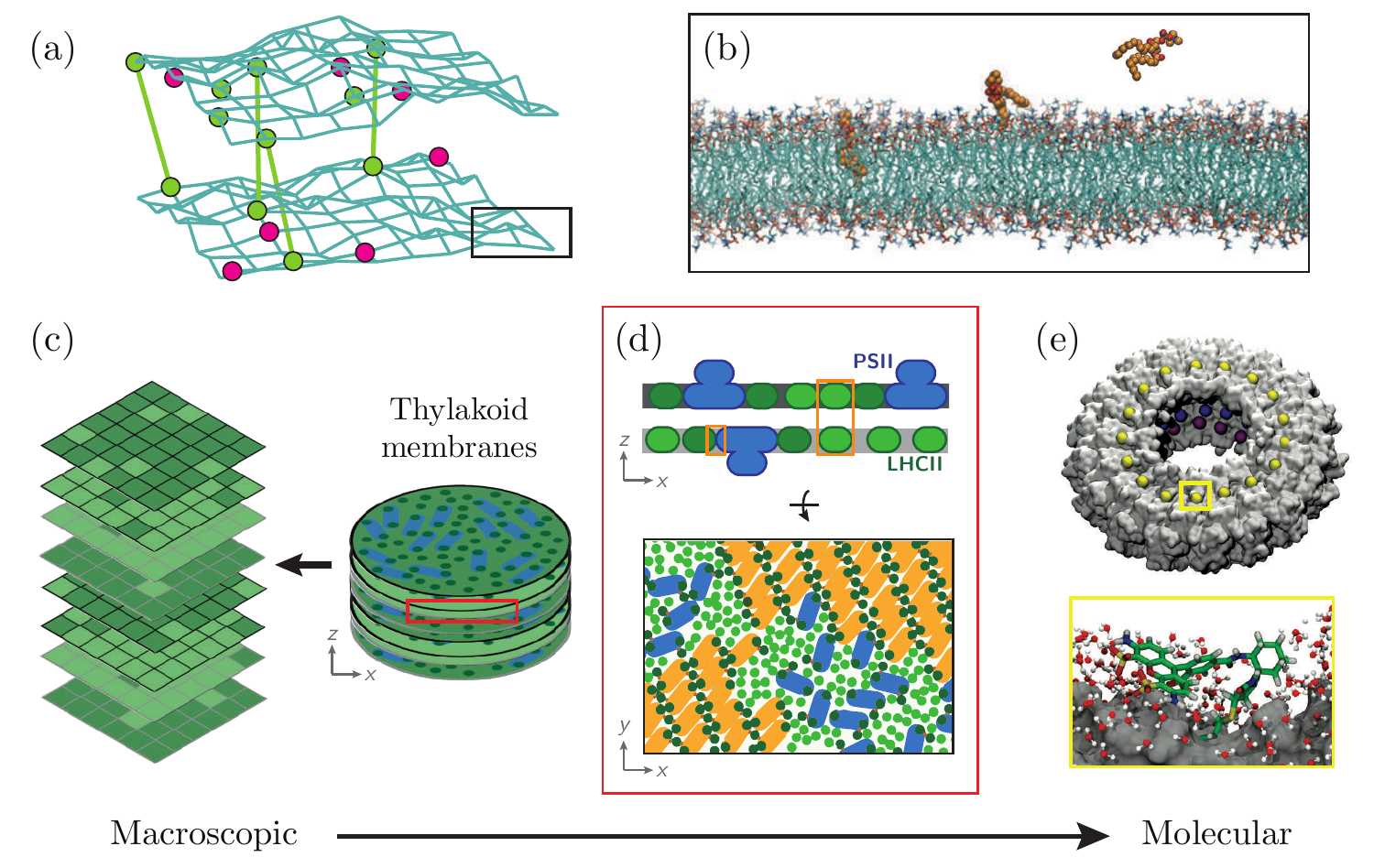} 
   \caption{Phill's work on biological membranes and the processes embedded within their milieu bridged macroscopic and molecular phenomena by combining simple coarse-grained models with atomistically detailed simulations: (a) Micron scale membrane interfaces were modeled as an elastic mesh decorated by proteins~\cite{schmid_size-dependent_2016}, whereas (b) the exchange of individual lipids between membranes necessitated atomistic simulation~\cite{rogers_breakage_2020,rogers_membrane_2021}. The influence of thylakoid's mesoscale vertical structure on protein organization was captured with (c) lattice models~\cite{rosnik_lattice_2020}, whereas organization within the plane was recapitulated with (d) a nanoscale coarse-grained model~\cite{schneider_coarse-grained_2014}. (e) Optimal molecular environments for positioning light-harvesting sites within protein scaffolds were investigated with atomistic models~\cite{noriega_manipulating_2015,delor_exploiting_2018}.}
   \label{fig:mem_photosyn}
\end{figure}

Lipid membranes not only exhibit large length scale fluctuations resembling that of elastic sheets (Fig.~\ref{fig:mem_photosyn}(a)) but also variations at the molecular scale (Fig.~\ref{fig:mem_photosyn}(b)). Understanding how phenomena at these disparate scales are coupled intrigued Phill and inspired him to devise novel theories and computational methods. For example, by solely accounting for hydrophobic forces of association and the requirement of high equatorial density, he and his group created a deceptively simple model that recapitulates the elasticity and fluidity of natural membranes \cite{pasqua_large-scale_2010}. In another example, Phill and his group formulated a theory to explain another set of experiments conducted by Dan Fletcher and coworkers that showed membrane curvature can be driven by protein--protein crowding \cite{stachowiak_membrane_2012}.

A membrane's physical properties can also be specifically altered through the exchange of individual lipids between membranes (Fig.~\ref{fig:mem_photosyn}(b)). A detailed understanding of lipid-exchange dynamics had remained elusive, in part due to discrepancies between experiments and previous molecular simulations. Phill often reminded his students of the perils of presupposing a reaction coordinate to investigate dynamical events and how a poor choice could obscure the rate-limiting free energy barrier. So, Phill and his group took an alternative approach to that of previous computational studies and harvested natural, unbiased trajectories of lipid insertion. Using committor analysis (aka the ``Geissler Test'' in Fig.~\ref{fig:water}(a)), they revealed that the breakage of hydrophobic lipid--membrane contacts limits the rate of passive lipid transport, resolving the earlier discrepancies between experiment and simulation \cite{rogers_breakage_2020}. Importantly, knowledge of the reaction coordinate enabled the construction of a Smoluchowski equation for the rate of lipid exchange to model length and time scales probed in experiment \cite{rogers_breakage_2020}; allowed for systematic investigation of the membrane physicochemical properties that impact lipid transport rates \cite{rogers_membrane_2021}; and provided a foundation to understand the catalytic function of lipid-transfer proteins \cite{rogers_ceramide_2022}.

Working on lipid membranes naturally sparked Phill's interest in the biological processes embedded within their milieu. Many processes, ranging from cell-cell communication to photosynthesis, require specific arrangements of membrane proteins. Phill, often in close collaboration with experimentalists, sought to uncover general physical principles responsible for such spatial organizations. For example, through Monte Carlo simulations of a deformable, fluid membrane interface decorated with proteins defined by their heights and binding potentials (Fig.~\ref{fig:mem_photosyn}(a)), Phill and his group recapitulated the size-dependent segregation of proteins at membrane interfaces observed in reconstituted experiments conducted in Dan Fletcher's lab. Moreover, the simulations demonstrated how the interplay of protein height, lateral crowding, binding affinity, and thermal membrane height fluctuations collectively contribute to the formation of characteristic patterns of intracellular signaling \cite{schmid_size-dependent_2016}.

%Subsection: Photosynthetic systems
Motivated to uncover the physical driving forces underlying the organization of photosynthetic proteins in thylakoid membranes, Phill and his group devised minimal models amenable to thorough analytical and computational investigation (Fig.~\ref{fig:mem_photosyn}(c and d)). Their model of protein organization within appressed membranes of thylakoid disks included just two particle types, representing photosystem II (PSII) supercomplexes and light-harvesting complex II (LHCII) trimers, with simple shapes and short range inter-particle interactions, chosen based on structural studies. The elegant simplicity by which the model reproduced experimental observations, including extended PSII arrays that had eluded previous computational studies \cite{schneider_coarse-grained_2014}, has made it an ideal starting point for the development of more detailed models \cite{amarnath_multiscale_2016,bennett_energy_2018,wood_modeling_2020}. Furthermore, these simulations revealed the existence of a physiologically relevant phase transition between a disordered PSII-LHCII fluid and dense PSII-LHCII crystal \cite{schneider_coexistence_2013} (Fig.~\ref{fig:mem_photosyn}(d)). This led Phill and his group to explore if the thylakoid's mesoscale vertical structure modulates such phase behavior through a minimally detailed lattice model of stacked discs that captured the alternating attractive and repulsive forces acting between vertically aligned membranes. Combining computer simulations with mean-field analysis, they found that a modulated phase with long-range order would form under certain conditions~\cite{rosnik_lattice_2020} (Fig.~\ref{fig:mem_photosyn}(c)). Phill was keenly aware of the biological implications of phase transitions \cite{schneider_coarse-grained_2014}. In the context of photosynthesis, he highlighted how proximity to phase coexistence could facilitate significant collective reorganization to alter thylakoid function in response to subtle environmental changes \cite{schneider_coexistence_2013, rosnik_lattice_2020}.

In parallel to investigating natural photosynthetic proteins, Phill alongside experimental collaborators Matt Francis and Naomi Ginsberg found artificial light-harvesting systems fruitful for deconvolving how individual molecular components concertedly impact energy-transfer dynamics. Recapitulating experimental results, simple lattice models of self-assembling mosaic virus capsid proteins illustrated how they could be used as a scaffold to arrange chromophores in geometries optimal for energy transfer~\cite{miller_impact_2010,hamerlynck_static_2022}, and atomistic simulations elucidated how each chromophore's protein and solvent environment could be tuned to extend photoexcitation lifetimes~\cite{noriega_manipulating_2015,delor_exploiting_2018}. Such studies illustrate the tact with which Phill combined experimentally grounded coarse-grained simulations with atomistic models to provide holistic pictures of processes spanning disparate length and time scales.

%% file: selfassembly.tex
%SW: I edited the section with input (paragraphs, comments, suggestions, plus old presentations and reference letters) from Eran R, David R, JiYeon, Asaph, Michael, Grant, Ahmad, Katie. There were others who would have contributed had they not been traveling (and so I prefer to be as inclusive as possible when we list authors).

%Good comments & suggestions, thanks. I will try to go through this one more time.

\section{Self-assembly} 
\label{sec:selfassembly}

Phill was fascinated by nanoscale self-assembly, the spontaneous organization of small molecules, nanoparticles, or biological complexes into ordered structures~\cite{whitesides2002self}. Self-assembly was fertile ground for one of his favorite strategies, ``identifying the essential microscopic variables whose fluctuations cannot be ignored''. He explored the thermal fluctuations that drive Brownian motion and make self-assembly possible, with structural order emerging from thermal disorder. He thought about design strategies for reliably assembling collections of weakly interacting components in the face of thermal buffeting~\cite{whitelam_role_2009,grunwald_patterns_2013}. He created models to show how the outcome of assembly could depend crucially on thermal fluctuations of density, conformation, or solvent~\cite{rabani_drying-mediated_2003,ku_self-assembly_2011,whitelam_impact_2009}. He thought about driven and non-driven systems in similar terms, seeking to understand pattern formation in a unified way~\cite{grunwald_exploiting_2016,klymko_microscopic_2016}, and thought about how nonequilibrium controls could be imposed to direct assembly in simulation and experiment~\cite{rabani_drying-mediated_2003,ku_self-assembly_2011,grunwald_exploiting_2016}.

\begin{figure}[]
   \centering
   \includegraphics[width=\linewidth]{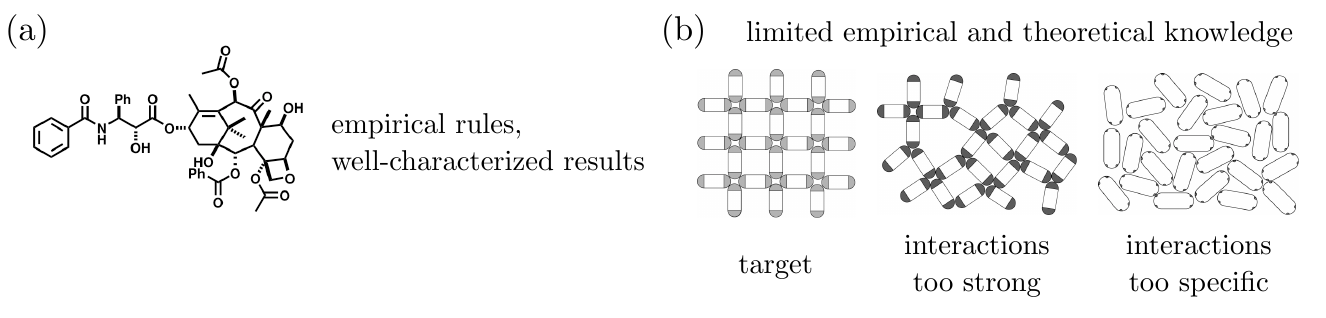} 
   \caption{Phill's approach to self-assembly was motivated by the gap in our understanding between (a) synthesis driven by covalent chemistry and (b) nanoscale organization controlled by weak interactions.}
   \label{fig1}
\end{figure}

Phill's work on self-assembly began with a desire to understand the dynamical pathways that result in the self-assembly of ordered structures, and to reveal guiding principles that allow for greater experimental control of these pathways. He approached this problem with an awareness of the contrast between our rudimentary understanding of assembly and our extensive empirical understanding of covalent chemistry and polymerization (Fig.~\ref{fig1}). In the early 2000s, the multi-step covalent synthesis of small molecules had become routine, driven by a detailed understanding of the reactions mediated by strong, highly directional covalent forces. The rules governing molecular self-assembly were less clear, and Phill focused his attention on the key role played by thermal fluctuations. These give rise to Brownian motion, the means by which nanoscale components encounter each other, but they also disrupt the weak, noncovalent forces by which nanoparticles associate. Consequently, intermolecular forces must achieve a balance: they must be strong and directional enough to stabilize a target structure, but weak enough to allow thermal fluctuations to disrupt non-optimal contacts and so correct errors. Phill's thinking was inspired in part by model studies done in David Chandler's group~\cite{hagan2006dynamic}.

Phill explored the general principles of assembly through specific case studies. Many of these were inspired by experiment, and were done in collaboration with experimentalists. These studies typically tested the hypothesis that the thermally mediated fluctuations of a few key microscopic variables dictated the essence of the self-assembly seen in the laboratory. Phill and his group would represent these variables within a simplified, statistical mechanical model of the laboratory experiment, and simulate the model on the computer. Often they would observe striking agreement between simulation and experiment, validating the hypothesis.

\begin{figure}[]
   \centering
   \includegraphics[width=0.8\linewidth]{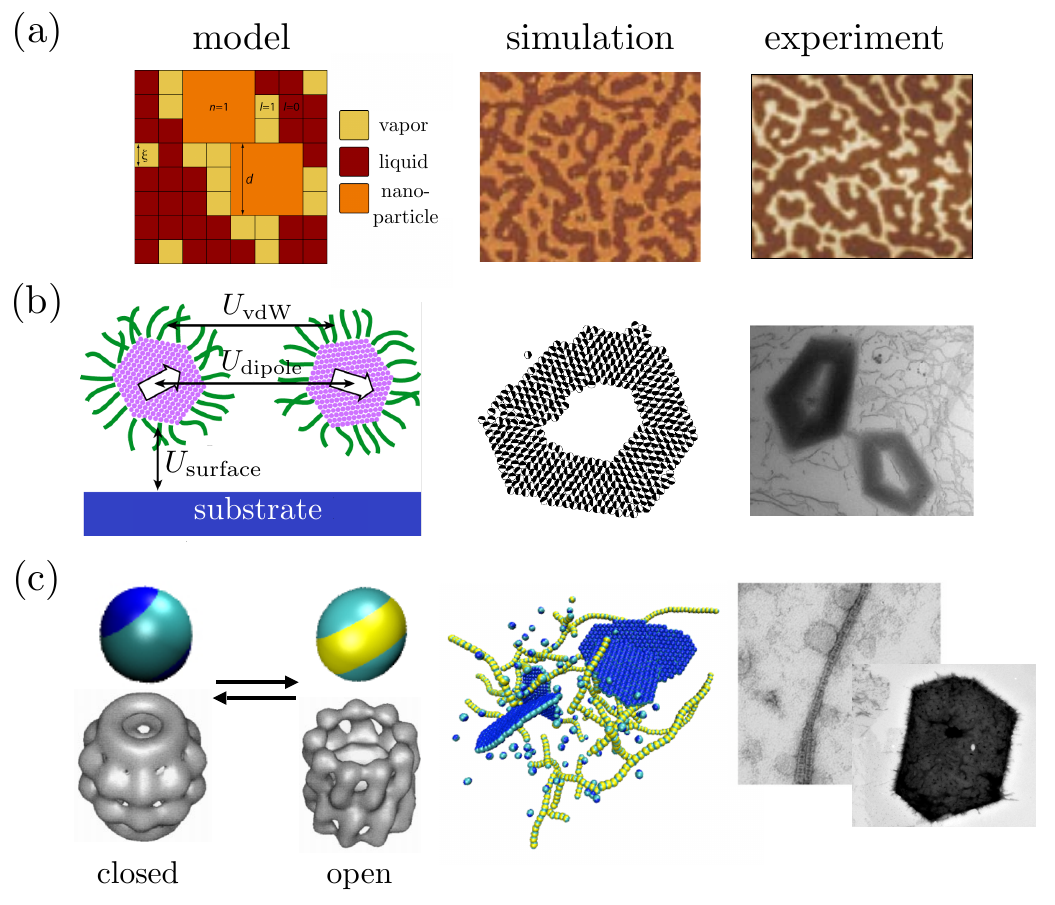} 
   \caption{Phill's approach to science often focused on identifying the microscopic variables whose fluctuations cannot be ignored. Doing so within models of self-assembly produced striking agreement between simulation and experiment for a variety of systems. Here the key fluctuations are those of (a) solvent, in a model of nanoparticles on a surface~\cite{rabani_drying-mediated_2003}; (b) nanoparticle density, in a model of magnetic nanoparticles~\cite{ku_self-assembly_2011}; and (c) nanoparticle conformation, in a model of protein complexes~\cite{whitelam_impact_2009}.}
   \label{fig2}
\end{figure}

Phill's first study of this kind was undertaken with longtime collaborators and friends Eran Rabani and David Reichman, and focused on understanding experiments in Louis Brus' laboratory~\cite{rabani_drying-mediated_2003}. There, nanoparticles self-assembled on a substrate as their solvent dried, forming a range of intricate structures. Prior work had identified the key role of solvent from a mean-field perspective~\cite{tang2002gas}, but Phill and his collaborators hypothesized the importance of solvent fluctuations. They designed a lattice model of nanoparticle self-assembly in which the solvent was represented in a coarse-grained but explicit way, using Ising-like degrees of freedom. In simulations, different rates of drying led either to spinodal-like or nucleation-like evaporation of the solvent, in each case inducing the formation of distinctly different self-assembled structures. These structures closely resembled the experimental assemblies (Fig.~\ref{fig2}(a)), validating the hypothesis and identifying a key means of control for this type of self-assembly. This perspective has had lasting impact, and has provided a unified view of a collection of experiments involving a broad range of specific materials~\cite{park_direct_2012,gordon2020automated}.

In a second study involving self-assembly mediated by drying, Phill and his group sought to explain the formation of hollow polygons formed by magnetic cobalt nanoparticles in Paul Alivisatos' laboratory. Similar magnetic nanoparticles had previously been seen to self-assemble into chains and loops of a single-nanoparticle width; the formation of hollow, faceted structures several nanoparticles wide was a puzzle. Phill and his group developed a coarse-grained model that described nanoparticles as dipolar spheres bearing short range van der Waals forces and long range dipolar forces~\cite{ku_self-assembly_2011}. Within the model, nanoparticle density fluctuations during drying led to the formation of nanoparticle aggregates. The long range, anisotropic nanoparticle interactions caused aggregates to adopt hollow, faceted shapes, strikingly similar to those seen in experiment (Fig.~\ref{fig2}(b)).

Phill turned his attention to experiments in which the key fluctuations were those of the nanoparticles themselves. Jonathan Trent and Chad Paavola at NASA Ames had conducted self-assembly experiments with protein complexes called rosettasomes. They found rosettasomes to self-assemble, under identical conditions, into filamentous structures and planar structures. Such polymorphism is peculiar because the kinetics of formation of one-dimensional structures is generally different to that of two- or three-dimensional structures; to have both self-assemble at the same time is unusual. Phill and his group hypothesized that this polymorphism resulted from the ability of the rosettasome to adopt different conformations as assembly proceeded~\cite{whitelam_impact_2009}. Simulations of patchy nanoparticles showed that conformational fluctuations could indeed drive polymorphism of one- and two-dimensional assemblies (Fig.~\ref{fig2}(c)).

Phill's work on self-assembly also confronted a computational issue: how can we efficiently simulate the self-assembly of components from solution? The most accurate classical approach, all-atom Newtonian dynamics with explicit solvent, is in general too expensive, requiring the evolution of millions of atoms over minutes or hours. An alternative approach is to represent the solvent implicitly, and evolve the nanoscale components using overdamped Langevin dynamics. This approach captures several important features of the all-atom approach, but underestimates the rate at which tightly bound collections of particles diffuse. Collective motion of this nature can cause kinetic trapping and enable hierarchical assembly, and so it is important to represent it accurately. To address this problem, Phill and his group developed a collective-move Monte Carlo algorithm for nanoscale components in solution~\cite{whitelam_avoiding_2007,whitelam_role_2009}. In effect a coarse-grained dynamics, the algorithm omits some fine details of real motion but captures two of its important features, moving nanoscale components locally according to the forces they experience and collectively in a way that approximates realistic diffusion. The algorithm has been used by other groups to study self-assembly, and incorporated into code for simulating DNA nanostructures~\cite{sengar2021primer}.

Phill continued to seek inspiration from experiment, and he and his group developed the theoretical underpinnings of one of the first experimental demonstrations of dense packings of polyhedral shapes on the nanoscale. Nanoparticle structures formed via solvent evaporation are often rendered imperfect by the nonequilibrium nature of the evaporation process, as Phill had shown previously, or by kinetic traps caused by strong particle interactions. However, experiments done in Peidong Yang's lab showed that gravitational sedimentation of polyhedral nanocrystals results in the self-assembly of highly ordered superlattices. These structures resemble the densest possible packings of mathematically perfect hard polyhedra. Phill and his group showed that the polymer chains present in solution are key for the self-assembly of uniform hard packings: they adsorb onto nanoparticle surfaces and provide a repulsive force that effectively cancels the attractive forces between nanoparticles, allowing them to behave like hard shapes~\cite{henzie_self-assembly_2012}. At high concentration the polymer chains induce depletion forces that lead to the formation of surprisingly complex, open lattices of polyhedra. Phill's rationalization of intriguing experimental results in terms of an interplay of driving forces and competing interactions is a hallmark of his work on self-assembly. He and his coworkers would later draw on similar ideas to understand nanoparticle surfactant assembly and jamming at the oil-water interface~\cite{chai_direct_2020}.
     
Studying biological systems provided inspiration for the design of synthetic ones. Biological assemblies are driven by patchy interactions whose geometry encodes the target structure. Typical nanoparticle interactions lack the specificity of biological components, and their assemblies are less complex. Phill was fascinated by the idea of creating experimental pathways to complex self-assembly, particularly without the need for sophisticated building blocks. He and his group demonstrated that assemblies with intricate spatial and compositional structures, of varying dimensionality, could be generated from a small number of simple spherical component types that assemble hierarchically into effective patchy nanoparticles~\cite{grunwald_patterns_2013}. The assembly strategy suggested in that paper has received much attention and has inspired experimentalists to build similarly patchy building blocks~\cite{porter2017directed}.

Phill continued to think about the nonequilibrium controls that can be used to direct assembly and pattern formation. Working with the group of George Whitesides, Phill's group demonstrated that the mechanical agitation of macroscopic particles leads to unexpected self-assembly behavior that cannot be explained by equilibrium fluctuations. Instead, the mechanical parameters of shaking devices induce mobility differences between particles that lead to effective attractive interactions~\cite{grunwald_exploiting_2016}. While the experiments were macroscopic, the paper showed that the same principles could be used to understand the driven self-assembly of microscopic particles in solution. Indeed, similar physics is seen in colloidal mixtures in which two species of particle are driven in opposite directions, forming lanes parallel to the direction of driving~\cite{vissers2011lane}. Simulations by others showed that lanes result from the enhanced diffusion of particles when surrounded by particles of the opposite species. Phill and his coworkers showed that such enhanced diffusion is a geometric effect that results from rectification of particles' Brownian fluctuations~\cite{klymko_microscopic_2016}. Simple scaling arguments reveal the dependence of this enhancement on the strength of the drive, providing guidance for control of the phenomenon.

Much of Phill's work on self-assembly focused on understanding the rules by which nanoscale components can avoid kinetic traps and self-assemble into thermodynamically stable structures. His fascination with biological assembly led him to examples in which, instead, the thermodynamic structure was not useful and the functional assemblies were kinetically trapped. Electron tomography studies of the $\beta$-carboxysome, a focus of David Savage’s lab at Berkeley, showed that it self-assembled into a surprisingly uniform icosahedral structure with a narrow distribution of sizes~\cite{iancu_organization_2010}. Phill found this observation fascinating because the carboxysome, unlike simple small viruses, consists of proteins assembling around a condensed cargo that could in principle grow without bound. He and his coworkers introduced a minimal model that showed that the equilibrium structure would indeed never consist of an encapsulated cargo of finite size~\cite{rotskoff_robust_2018}. They built a model that captured the essential mechanics and dynamics of carboxysome assembly, and showed that finite-size encapsulation was possible in the form of a kinetically trapped, nonequilibrium structure. Moreover, the kinetics of assembly could be tuned to produce structures of different sizes, with a dispersion controlled by the mismatch between the rate of growth of the carboxysome cargo and its protein shell.

\begin{figure}[]
   \centering
   \includegraphics[width=0.8\linewidth]{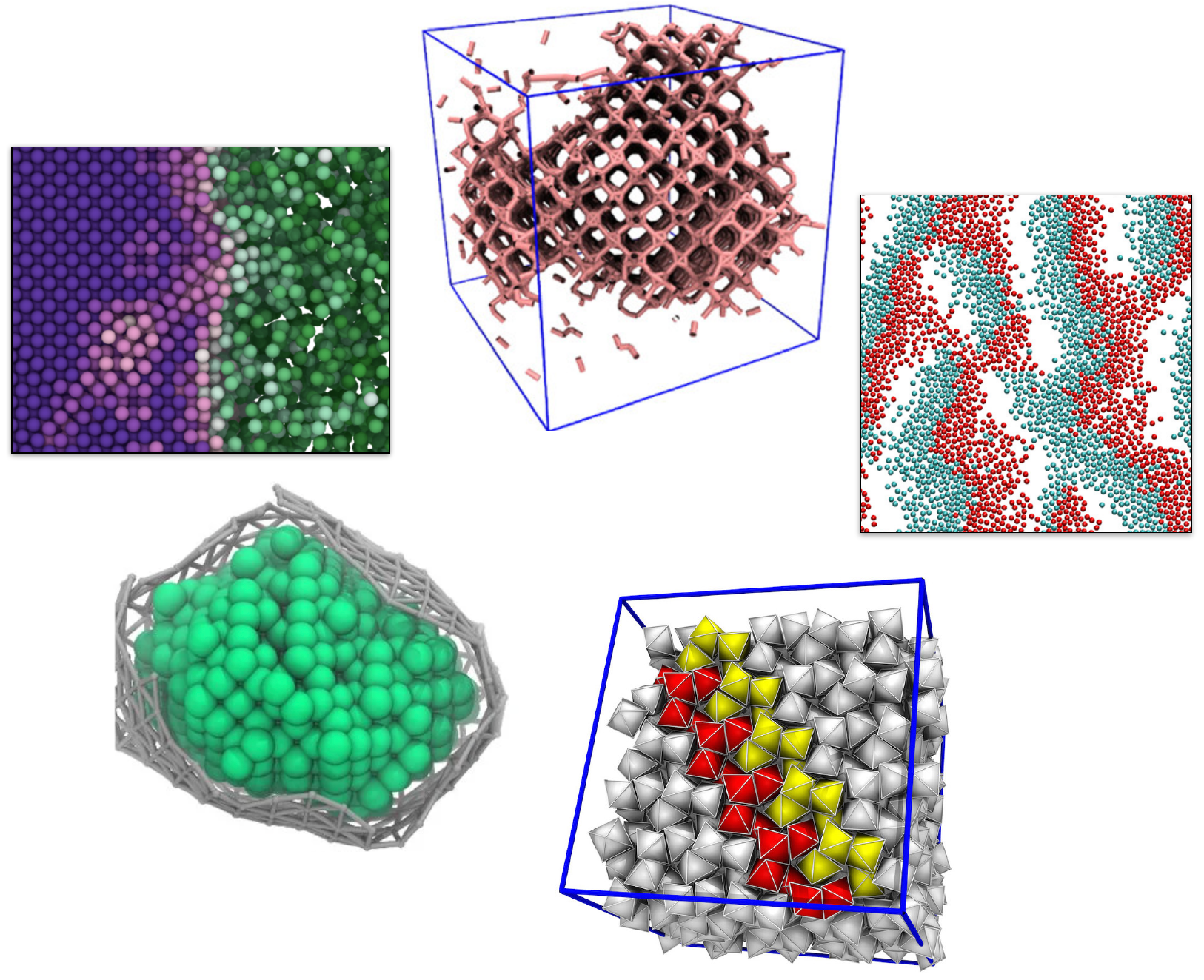} 
   \caption{Phill appreciated the beauty of self-assembly. Clockwise from top: models of emergent patchy particles~\cite{grunwald_patterns_2013}; shaken plastic beads~\cite{grunwald_exploiting_2016};  faceted nanocrystals~\cite{henzie_self-assembly_2012}; the carboxysome~\cite{rotskoff_robust_2018}; and active particles~\cite{omar2021phase}.}
   \label{fig3}
\end{figure}

Phill also studied the behavior of active particles. These are energy-consuming units such as bacteria that are capable of self-propulsion. Active particles can phase separate in the absence of attractive interactions, driven by a feedback effect whereby particles accumulate where they slow and slow where they accumulate~\cite{cates2015motility}. Phill and his coworkers demonstrated that self-assembling active systems bear a closer resemblance to self-assembling passive systems than previously appreciated~\cite{omar2021phase}. In particular, active particles in three dimensions can achieve three-phase coexistence of solid, liquid, and gas, similar to the triple point of a substance such as water. Three-dimensional active systems also exhibit metastable liquid-gas coexistence above a triple point, and Phill and coworkers used tools from large-deviation theory to argue that such metastability is a generic feature of equilibrium and nonequilibrium systems. This work recalls Phill's ability to identify common physics in seemingly disparate systems, providing insight into self-assembly and nonequilibrium statistical mechanics more broadly.

 Phill's work on self-assembly focused on the fundamentals and was mindful of the applications. He sought to identify the basic physics of molecular scale organization, motivated by an understanding of the importance of self-assembly to biology and materials science. But Phill also appreciated the intrinsic beauty of self-assembly, and encouraged his group members to highlight this beauty in their work (Fig.~\ref{fig3}).

%% file: nanomaterials.tex
\section{Nanomaterials}
\label{sec:nanomaterials}
% Authors: Layne Frechette , Michael Gr\"unwald, Asaph Widmer-Cooper
%Contact info:
% laynefrechette@brandeis.edu

\begin{figure}[]
   \centering
   \includegraphics[width=\linewidth]{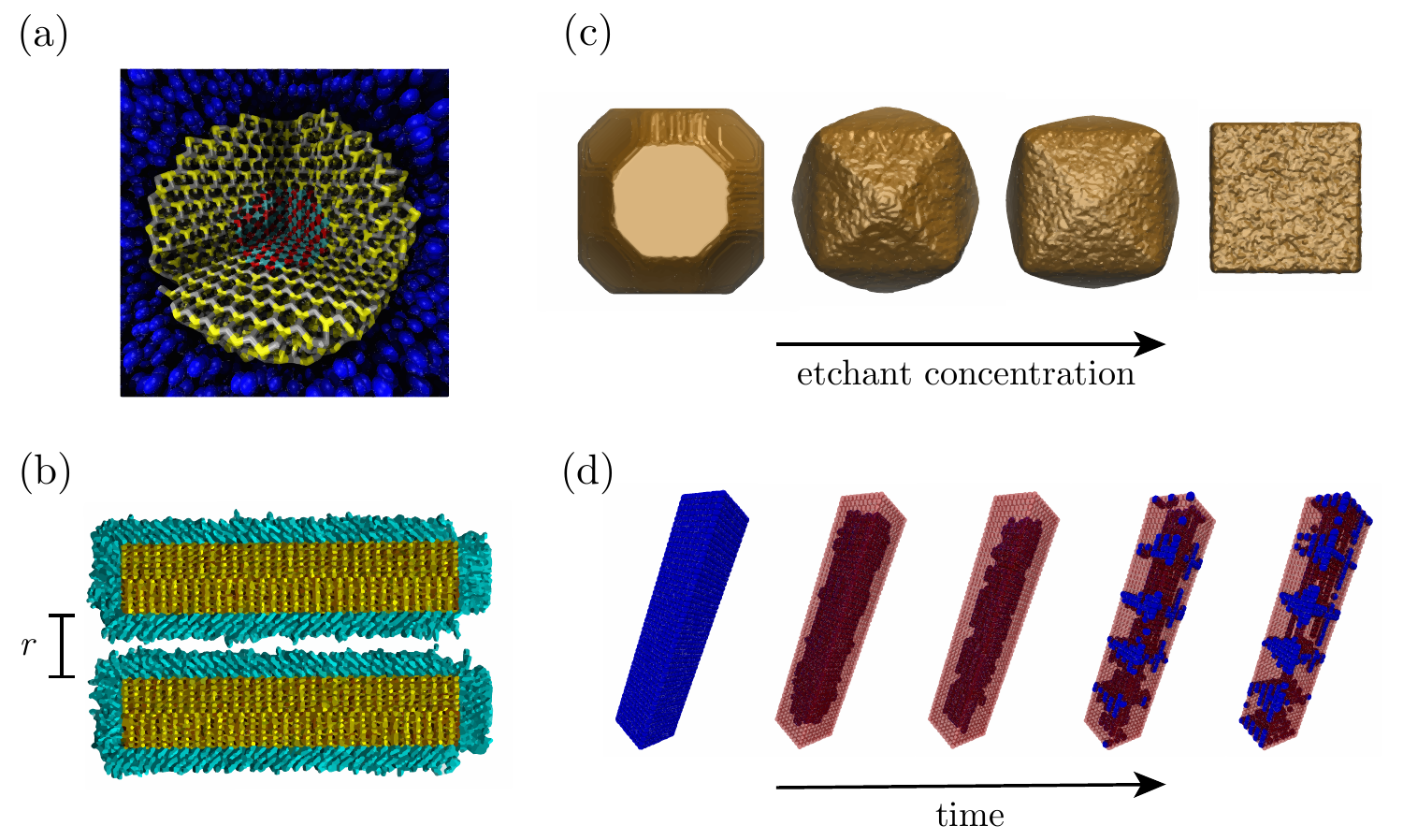} 
   \caption{Phill explored the boundary between the macroscopic and the microscopic in his work on nanomaterials. He created molecularly detailed models to elucidate (a) pressure-induced phase transitions in core/shell nanocrystals~\cite{grunwald_metastability_2013} and (b) interactions between nanorods mediated by organic ligands and solvent~\cite{widmer-cooper_orientational_2014}. He also devised minimal models which captured the emergent dynamics of (c) nanocrystal etching ~\cite{ye_single-particle_2016} and (d) cation-exchange reactions~\cite{frechette_elastic_2021}.}
   \label{fig:nanomaterials}
\end{figure}

Phill was fascinated by the chemistry of nanomaterials, which he described as lying on ``a scale between macroscopic and microscopic where things work differently.” It is here where ``more than a few, less than a lot” of molecules contribute to emergent behavior and material properties that are scientifically intriguing and technologically promising~\cite{berkeleychemistryliveHowNanoscienceChanges2021}. Driven by Phill's deep understanding of both macroscopic thermodynamics and microscopic fluctuations, his group contributed to a range of nanoscale problems, including structural and compositional transformations of nanocrystals~\cite{grunwald_metastability_2013,ye_single-particle_2016,hauwiller_unraveling_2018,frechette_consequences_2019,frechette_origin_2020,frechette_elastic_2021}, formation of nanomaterials~\cite{widmer-cooper_orientational_2014,widmer-cooper_ligand-mediated_2016}, and new computational methods and models to study phenomena on the nanoscale~\cite{grunwald_transferable_2012,grunwald_efficient_2007,grunwald_precision_2008}. 
Phill's work on nanomaterials was profoundly impacted by Paul Alivisatos' research group: their precise physical measurements and high-resolution characterization of nanomaterial transformations offered an ideal complement to Phill's approach to understanding nanoscale systems.  

One of Phill's earliest investigations of nanomaterials was inspired by high-pressure experiments performed in the Alivisatos group, which probed solid-solid phase transformations in nanocrystals. These experiments showed strongly size-dependent transformation hysteresis, and suggested the possibility of kinetically trapping nanocrystals in metastable crystal structures. To investigate the rare nucleation events at the heart of these nanocrystal transformations, Phill and his collaborators developed a transition path sampling algorithm that used an ideal gas to apply pressure on single nanocrystals \cite{grunwald_efficient_2007}.
Simulations using this algorithm revealed the microscopic mechanism underlying nanocrystal transformations and rationalized the experimentally observed transition kinetics. In related work, Phill’s group in collaboration with the Alivisatos lab investigated the thermodynamics and kinetics of structural transformations in core-shell nanocrystals \cite{grunwald_transferable_2012,grunwald_metastability_2013}. They showed that by combining structurally related materials with different transition pressures in a core-shell geometry, new crystal structures, inaccessible in the bulk, could be kinetically trapped under high pressure and stabilized by tuning the thickness of the nanocrystal shell. Phill’s work on nanocrystal transformations is a prominent example of how his group combined method development and model development to address intriguing experimental questions at the forefront of nanoscience.  

Inspired by work with the Alivisatos group on the self-assembly of colloidal nanorods~\cite{baker_device-scale_2010}, and echoing concurrent work on the self-assembly of rosettasomes~\cite{whitelam_impact_2009} described in Sec.~\ref{sec:selfassembly}, Phill became interested in the possibility that the self-assembly of nanoparticles could be influenced by structural changes within their ligand shells. He and his group used molecular dynamics simulations to study how passivating ligands order on nanorods in solution and how that affects the interaction between the particles. This work predicted that even in relatively good solvents, the ligands could transition from a mobile disordered state to a less mobile one where the ligands were packed together and orientationally ordered with one another, simultaneously changing the rod-rod interaction from repulsive to attractive~\cite{widmer-cooper_orientational_2014}. Subsequent work showed that, as a consequence, interactions between nanoscale surfaces can depend sensitively and nonlinearly on temperature, facet dimensions and ligand coverage~\cite{widmer-cooper_ligand-mediated_2016}. In later years, Phill continued to explore this problem with his group, working to develop and parameterise a simple phase-field model~\cite{satishMappingPhaseDiagram} that could be used to study the interplay between ligand ordering and nanoparticle assembly.

When Eran Rabani moved to a faculty position at Berkeley, he and Phill became interested in explaining other experiments from the Alivisatos lab in which nanocrystals dissolved, or were “etched,” in an oxidant-rich solution. On their way to complete dissolution, these nanocrystals adopted different shapes depending on the concentration of etchant. To understand these shape transformations, Phill and coworkers took inspiration from their previous work on evaporation-induced nanoparticle assembly (described in Sec.~\ref{sec:selfassembly}) and employed a lattice model to describe nanocrystal dissolution. Here the process of etching was represented simply by a chemical potential difference driving the removal of occupied sites at the nanocrystal surface, and coordination number determined the rates of material removal at different surface locations. Consistent with experiments, kinetic Monte Carlo simulations of the lattice model exhibited different shape transformations as the driving force was varied. A detailed analysis of the simulated etching trajectories revealed that different regimes of shape transformation corresponded to which types of surface atoms etched at an appreciable rate. The driving force could be tuned, for example, to a value such that surface sites with coordination number 6 or lower were all removed with approximately the same rate, while those with coordination number 7 or higher were removed much more slowly. This etching dynamics promoted the formation of different geometrical facets on the nanocrystal surface at different values of the driving force (a mechanism dubbed “step-recession.”) The resulting shape transformations matched closely with those observed in experiments \cite{ye_single-particle_2016,hauwiller_unraveling_2018}.

Phill had also long been intrigued by another set of experiments from the Alivisatos lab in which ions of one species are replaced by those of another in a nanocrystal \cite{sonCationExchangeReactions2004a,robinsonSpontaneousSuperlatticeFormation2007}. These cation-exchange experiments produce a diverse array of heterostructures on the way to complete replacement. Initial attempts to understand these reactions via computer simulation using detailed molecular models suffered from small trajectory sample sizes and an inability to access the relevant time scales. Taking a different tack, Phill worked with long-time friend and collaborator Christoph Dellago to develop a simplified lattice model that focused on a key feature of cation-exchange reactions: the elastic strain that attends a mixture of different-sized ions. Computer simulations of the model yielded exchange trajectories featuring heterostructured intermediates, including striped nanocrystals resembling those seen in experiments \cite{frechette_elastic_2021}. Informed by the bulk equilibrium behavior of the elastic model---for which he and his group developed successful theories \cite{frechette_consequences_2019,frechette_origin_2020}---Phill explained the origin of these structures. The strong, nonequilibrium driving force for cation exchange creates effective, transient boundary conditions, mimicking those of a bulk system at equilibrium in which spatially modulated structures are thermodynamically stable. Through their investigations of cation exchange, Phill and his group highlighted the rich pattern formation that arises from the interplay of kinetics, geometry, and elasticity at the nanoscale.

%% file: methods.tex
\section{Minimal models and methods for probing complex phenomena} \label{sec:methods}

% Authors: Grant Rotskoff 
% Contributed text from: Christoph Dellago, Suri Vaikuntanathan

%characterized by widely separated time scales. The efforts of the Chandler group led 

Phill often described his work as “curiosity-driven”, and he did not limit himself to any particular method or scale. As we have emphasized in other sections, Phill appreciated the importance of dynamical trajectories and considered them revelatory for the physical processes he studied. When Phill joined the research group of David Chandler as a Ph.D. student in the fall of 1996, the group was intensively working on a new computational approach to study rare events---such as phase transitions, chemical reactions, and biomolecular reorganizations---characterized by widely separated time scales. 
Phill participated in Chandler group brainstorming sessions as transition path sampling (TPS) was being developed.
TPS is a Monte Carlo method in which ``moves'' in trajectory space are used to generate an ensemble of reactive trajectories. 
Importantly, unlike most rare-event methods, the procedure requires no prior knowledge of the reaction mechanism in the form of a transition state or a reaction coordinate ~\cite{bolhuis_transition_2000,dellago_transition_2002,bolhuis_transition_2002,ferrario_transition_2006}. 
With characteristic insight, Phill pointed out how to leverage the analogy between time correlation functions and the reversible work required to transform ensembles of trajectories~\cite{dellago_monte_2003,geissler_equilibrium_2004}.

By harvesting ensembles of reactive trajectories with TPS, mechanistic details could be explored, all while preserving the full complexity of every fluctuation. While Phill was able to derive significant physical insight from collections of reactive trajectories, actually collecting these trajectories required significant methodological innovation to make TPS tractable. For example, to study the microscopic mechanism for proton transfer in the protonated water trimer \cite{geissler_chemical_1999,geissler_ab_2000,geissler_potential_2000}, Phill came up with a smart way to perturb the points from which a trial trajectory is launched so that linear and angular momenta were conserved. 
Phill's simulations showed that the proton transfer is driven by the rearrangement of the oxygen ring rather than by the proton coordinate, a non-obvious reaction coordinate, as discussed in Sec.~\ref{sec:water}. 
To identify the relevant degrees of freedom for this process, Phill invented an ingenious way to test candidate reaction coordinates based on the calculation of committor distributions, as discussed in Sec.~\ref{sec:water} and Fig.~\ref{fig:water} (a).

\begin{figure}
    \centering
    \includegraphics[width=\linewidth]{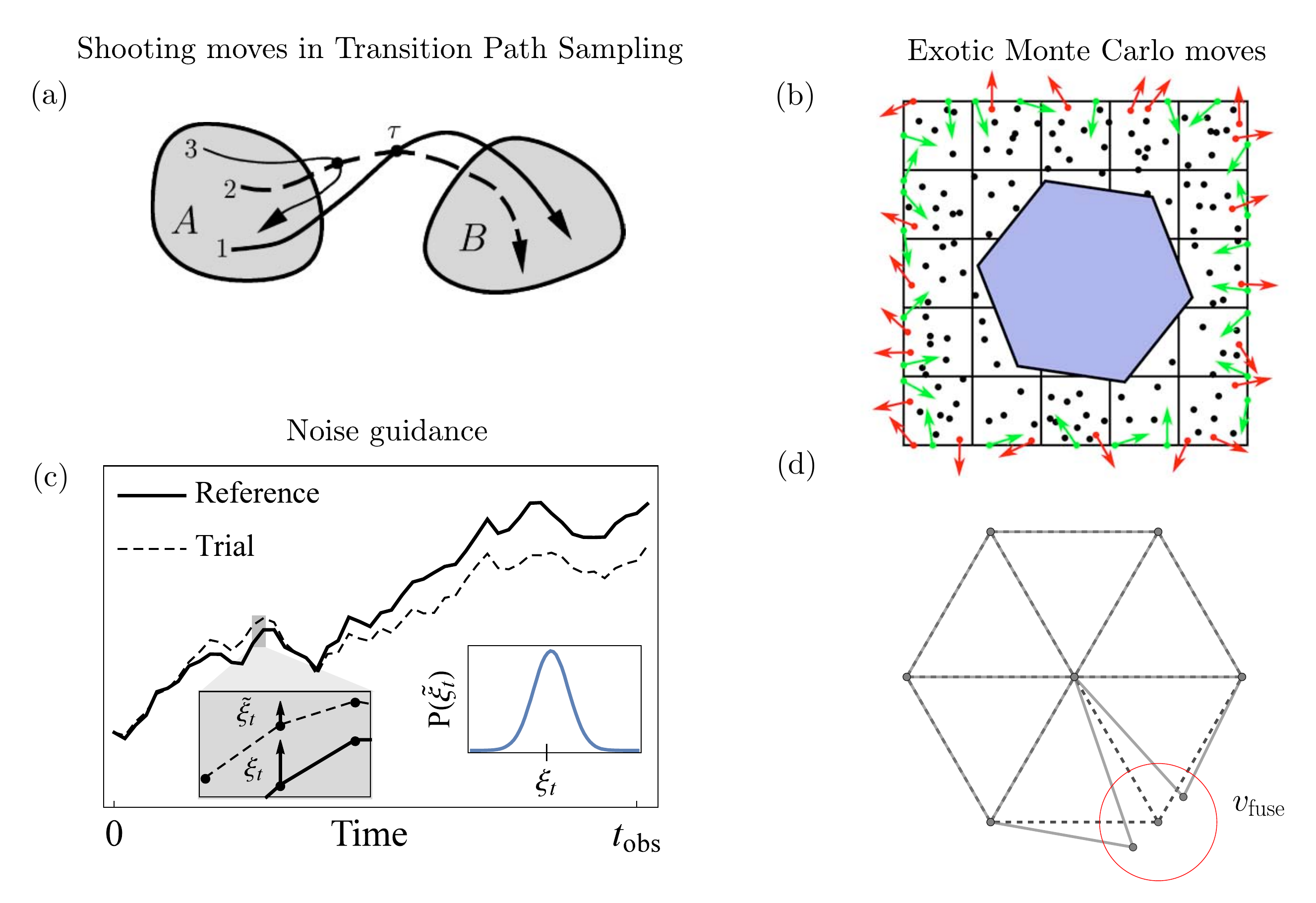}
    \caption{Phill's work on method development was imbued a philosophy of doing things ``the right way'', with precision and rigor: (a) schematic depiction of a shooting move in TPS from Ref.~\cite{bolhuis_transition_2002}; (b) and (d) schematic depictions of complex MCMC proposal moves from Refs.~\cite{grunwald_efficient_2007,rotskoff_robust_2018}, respectively; (c) an example correlated trajectory generated with the noise guidance algorithm from Ref.~\cite{gingrich_preserving_2015}.  }
    \label{fig:methods}
\end{figure}

Transition path sampling became a core tool within the Geissler group and a distinctive part of its philosophy.
Indeed, the development of trajectory sampling algorithms remained a focus throughout Phill's career, always combining imaginative ideas with mathematical rigor. For instance, he developed an efficient TPS scheme to study pressure-induced phase transitions in nanoparticles immersed in an ideal-gas pressure bath \cite{grunwald_efficient_2007}. In other work, Phill invented the method of precision shooting based on the linearized time evolution of small perturbations to control the acceptance rate in TPS simulations of long diffusive processes \cite{grunwald_precision_2008}. He returned to this problem later and designed ways to control the correlations between pathways by applying noise guidance to the generation of trajectories \cite{gingrich_preserving_2015}. More recently, Phill helped to apply machine learning to enhance TPS simulations by generating uncorrelated shooting points with normalizing flows \cite{falkner_ML_2022, coretti_ML_2022}. 
Phill always grounded his many creative contributions to the path sampling literature in the fundamentals of statistical mechanics, carefully deriving new methods from basic principles.    

Phill would often remark that he found the simplicity of maintaining detailed balance in transition path sampling beautiful, because the algorithm proscribes that a trajectory is accepted if, and only if, it is reactive. Acceptance criteria for complicated Markov Chain Monte Carlo (MCMC) moves can be subtle and many practitioners neglect the crucial “generation probabilities” that arise when a move is not statistically reversible. The simplest MCMC proposals, however, lead to slow relaxation and were often inappropriate for the physical systems that Phill studied. Moreover, Phill relished in working through complex MCMC procedures to obtain the correct acceptance probabilities. Exotic moves can be found tucked away in many of his papers~\cite{grunwald_efficient_2007,rotskoff_robust_2018}, and a handful are dedicated to establishing detailed balance for sufficiently complicated move
% not sure what the intent was here -- I would say "and some of his papers focus on the development of algorithms for evolving nanoscale systems."
% GR: the point was to highlight the fact that there are a few papers that are solely about DB in MCMC, Michael's paper on the barostat is about establishing detailed balance, but they show it doesn't change result that they had previously obtained
sets~\cite{whitelam_avoiding_2007,whitelam_role_2009,pronk_faster_2009}. 
%added virtual-move ref
%added Sander paper
Ensuring that everything was handled properly, even when it did not necessarily make a difference in typical simulations~\cite{grunwald_efficient_2007},
%preceding comment is a bit mysterious
embodies the careful method development that he engaged in.

% I think the main reason was simply that he loved universality
% I think it would be nice to mention Sarah Keller's work on bilayer phase behavior here too
% Phill valued path sampling because it captured the essential fluctuations that drove complex chemical processes. 
% However, he also had a deep appreciation for problems where the essential fluctuations could be described universally.
% Often these models were parameterized using data from experiments or atomistic simulations or were decorated with Gaussian field fluctuations that endowed them quantitative accuracy~\cite{geissler_importance_2000,vaikuntanathan_necessity_2016}. He also purpose built models for specific experimental systems, carefully designing them with a principled minimalism, attempting to retain only the fluctuations necessary to recapitulate a particular system. This philosophy is evident in his work on biophysical systems~\cite{whitelam_stretching_2008,whitelam_there_2008,gin_limited_2009,schneider_coexistence_2013,schneider_coarse-grained_2014,rosnik_lattice_2020,rotskoff_robust_2018,hohlfeld_communication_2014}, 
% as described in Sec.~\ref{sec:biophysics}. The “reductionist” philosophy and reconstitution experiments of his close collaborator Daniel Fletcher were particularly amenable to the minimalist approach. 

% Moving beyond the conventional applications of path sampling, Phill and his group also
% %suggest removing also
% applied these ideas to study of non-equilibrium processes. 
Mapping complicated phase behavior onto a lattice model was a particular passion of Phill’s, and Ising models appeared in his work on hydrophobic solvation~\cite{vaikuntanathan_putting_2014,vaikuntanathan_necessity_2016}, nonequilibrium solvation~\cite{geissler_importance_2000}, drying-mediated self-assembly~\cite{rabani_drying-mediated_2003}, cation exchange in nanocrystals~\cite{frechette_consequences_2019,frechette_origin_2020,frechette_elastic_2021}, and thylakoid membranes~\cite{rosnik_lattice_2020}.
Minimal models also provided Phill and his group with a lens to examine the dynamics of molecular systems driven away from equilibrium.
Nonequilibrium biological processes are often characterized by dynamical heterogeneity. This heterogeneity is apparent in a range of processes over many scales, from dynamical instability observed in microtubule growth to heterogeneity in cell growth rates. The latter is thought to enable a mode of antibiotic resistance in certain bacterial cells as slowly growing cells can have a higher probability of survival in the presence of antibiotics. Dynamical heterogeneity implies that these cells can then switch to the fast growth rate mode when conditions are more favorable~\cite{vaikuntanathan_dynamic_2014, gingrich_heterogeneity-induced_2014}. Phill and co-workers used the statistical mechanics apparatus developed in the context of path sampling to understand the basis for such phenomenology. By focusing on the so-called large deviation rate function, which plays a role formally analogous to that played by a free energy in equilibrium statistical mechanics, they revealed how dynamical heterogeneity and dynamical phase transitions can emerge due to the presence of seemingly minor heterogeneities in the kinetic rates. 
This work resulted in a minimal but analytically solvable model for dynamical phase transitions and heterogeneity. 
Phill and his group later adapted these ideas to probe efficiency fluctuations in a minimal model of a nanoscale Carnot cycle~\cite{gingrich_efficiency_2014}. Phill had a deep understanding of nonequilibrium fluctuation theorems and mapped out the implications of an asymmetric external driving protocol for the statistics of the fluctuating efficiency in this nanomachine.